\documentclass[11pt]{article}
\usepackage{graphicx}

\usepackage{amsmath}
\usepackage{amsfonts}

\def\a{\alpha}
\def\b{\beta}

\def\RR{{\rm I\kern-.17em R}}
\def\NN{{\rm I\kern-.17em N}}

\newtheorem{theorem}{Theorem}

\newtheorem{proposition}{Proposition}

\usepackage{epsf}
\usepackage{psfrag}
\usepackage{graphics}
\begin{document}

\title{On the Asymptotic Stability of De-Sitter Spacetime: a non-linear perturbative approach}

\author{Filipe C. Mena\footnote{email: fmena@math.uminho.pt}\\\\
{\em{Dep. Matem\'atica, Universidade do Minho, 4710-057 Braga, Portugal}}\\\vspace{6pt}}

\maketitle
\begin{abstract}
We derive evolution and constraint equations for second order perturbations of flat dust homogeneous and isotropic solutions to the Einstein field equations using all scalar, vector and tensor perturbation modes. We show that the perturbations decay asymptotically in time and that the solutions converge to the De-Sitter solution. By induction, this result is valid for perturbations of arbitrary order. This is in agreement with the cosmic no-hair conjecture of Gibbons and Hawking.
\\\\
keywords: General Relativity; Cosmic no-hair Conjecture; Non-linear Perturbations; Cosmological Attractor; Stability
\\\\
classcode: 83F05 ; 83C30; 53Z05
\end{abstract}
\section{Introduction}
A long standing problem in Mathematical General Relativity is the proof of the Gibbons-Hawking cosmological attractor conjecture, best known as the cosmic no-hair conjecture. This conjecture is the statement 
that generic expanding cosmological solutions to the Einstein Field Equations (EFEs) with a positive cosmological constant $\Lambda$ tend asymptotically in time to the De-Sitter solution \cite{GW}. Wald provided a proof of this conjecture within the class of spatially
homogeneous solutions \cite{Wald} while other authors, including Stein-Schabes and collaborators  \cite{BSS,JSS}, have used particular classes of inhomogeneous exact solutions with symmetries. So, it is important to consider more general non-symmetric cases and this can be achieved by using perturbation theory. An interesting result along this line was derived in \cite{Reula} where the author shows, subject to conditions on a symmetric hyperbolic system, the exponential decay of linear and non-linear perturbations around flat homogeneous and isotropic cosmologies. Other results, which do not necessarily imply an exponential decay, were obtained for linear metric perturbations around homogeneous and isotropic backgrounds by \cite{Starobinski, Boucher-Gibbons, Rendall}. 

A second order perturbative approach 
was followed by \cite{MBT} using scalar perturbations modes at first order which then coupled to source second order scalar and tensor perturbations. In particular, it has been shown that those first and second order perturbations decay and that the respective perturbed solution to the EFEs approaches the De-Sitter solution asymptotically in time \cite{MBT}. 

In this note, we generalize the results of \cite{MBT} in three ways: firstly, by including vector and tensor perturbations at first order (besides the scalar modes of \cite{MBT}). Secondly, by considering vector perturbations at second order, so that all scalar, vector and tensor modes are included at both first and second orders. Thirdly, by considering all those perturbations at any order.  
To do that, we shall first derive the second order perturbation equations around flat dust homogeneous and isotropic backgrounds by generalising the work of \cite{Mata}. We shall then estimate the asymptotic dynamics of the solutions and show that they decay in time. This will enable to prove that the perturbed spacetime asymptotes to De-Sitter. Finally, we shall show that, by induction, this result is valid for perturbations of any order. 

We use the following indice notation: small greek $\a, \b, ... =0, 1, 2, 3$ and small latin $a, b, ...= 1, 2, 3$.  
\\

We shall consider solutions to the EFEs such that $T_{\a\b}=\rho u_\a u_\b$, where $\rho$ is the mass density and $u_\a$ the 4-velocity of the fluid. We denote by $\Sigma$ the space orthogonal to $u$ on spacetime $(M,g)$.
Following \cite{Mata}, we use comoving euclidean coordinates, say $\{t,x^1,x^2, x^3\}$, to write the components of $u$ as $u^\a=(1,0,0,0)$ and, in this basis, the metric on $M$ can be written as:
\begin{equation}
\label{dust}
g=-dt^2+h_{ab}dx^a dx^b,
\end{equation}
where $h_{ab}$ is a function of all coordinates.
The components of the extrinsic curvature on $\Sigma$ coincide with the expansion tensor  (see \cite{Ehlers} for definitions about the 1+3 formalism)  and can be written as 
$$\theta^\a_{~\b}=\nabla_\b u^\a=\frac{1}{2}h^{\a\gamma}\dot{h}_{\gamma\b},$$
where the dot denotes differentiation with respect to $t$. 
In this case, the EFEs give (see also \cite{Mata}):
\begin{eqnarray}
\label{evolutiona}
&&\dot{\theta}_{\a\b}+\theta\theta_{\a\b}+R^*_{\a\b}=
\frac{1}{2}\rho\delta_{\a\b}+\Lambda\delta_{\a\b},\\
\label{Ray}
&&\dot{\theta}+\theta^{\a\b}\theta_{\a\b}+\frac{1}{2}\rho=\Lambda, \\
\label{mra1}
&&\nabla_\a\theta^\a_{~\b}=\nabla_\b\theta\\
\label{energy}
&&\theta^2-\theta^{\a\b}\theta_{\a\b}+R^*=2(\rho+\Lambda),
\end{eqnarray}
where we have denoted the components of the Ricci tensor on $\Sigma$ as $R^*_{\a\b}$ and  the Ricci scalar on $\Sigma$ as $R^*$.

\section{Perturbations around FLRW flat backgrounds}
In this section, we consider small perturbations of arbitrary order $n\in \NN$ on a flat Friedmann-Lema\^itre-Robertson-Walker (FLRW) background spacetime $(M_{\scriptscriptstyle \rm B}, g_{\scriptscriptstyle \rm B})$ with $\Lambda>0$ and $T_{\a\b}=\rho u_\a u_\b$. Following \cite{Mata}, we take the conformally rescaled FLRW flat metric $g_{\scriptscriptstyle \rm B}$
\begin{equation}
\label{de-sitter}
g_{\scriptscriptstyle \rm B}\equiv g^{(0)}_{\a\b}=a^2(\tau)(-d\tau^2+\delta_{ab}dx^a dx^b),
\end{equation}
where $\tau$ is the conformal time defined by $dt=a d\tau$ and $a$ is the scale factor. We consider perturbations in the form \cite{Mata}
\begin{equation}
\label{pert-metric}
g_{\alpha\beta}=g_{\a\b}^{(0)}+\sum_{n=1}^{+\infty} \frac{1}{n!}g_{\alpha\beta}^{(n)}
\end{equation}
with
\begin{eqnarray}
\label{bn1}
g_{00}&=&-a(\tau)^2\left(1+2\sum_{n=1}^{\infty}\frac{1}{n!}\psi^{(n)}\right)\\
\label{bn2}
g_{0a}&=&a(\tau)^2\sum_{n=1}^{\infty}\frac{1}{n!}V^{(n)}_a\\
\label{bn3}
g_{ab}&=&a(\tau)^2\left[\left(1-2\sum_{n=1}^{\infty}\frac{1}{n!}\phi^{(n)}\right)
\delta_{ab}+
\sum_{n=1}^{\infty}\frac{1}{n!}\chi^{(n)}_{ab}\right]
\end{eqnarray}
where $V^{(n)}_a$ can be decomposed as
\begin{equation}
\label{vecy}
V^{(n)}_{a}=\nabla_a \Phi^{(n)}+W^{(n)}_a,
\end{equation}
with $\nabla^a W^{(n)}_a=0$.
Now, for this case, as was shown by Stewart \cite{Stewart} (following works of 
Choquet-Bruhat, Fischer \& Marsden \cite{Choquet-etal} and D'Eath \cite{D'Eath}):
\begin{theorem} (Stewart) Any symmetric tensor $\chi^{(n)}_{\a\b}$ defined on three-surfaces $\Sigma$ of $M_{\scriptscriptstyle \rm B}$  
can be decomposed into three parts
\begin{equation} \label{decomp}
\chi^{(n)}_{ab}=(\nabla_a \nabla_b -\frac{1}{3}\delta_{ab}\Delta)\chi^{(n)}+2\nabla_{(a}Z^{(n)}_{b)}+\pi^{(n)}_{ab},
\end{equation}
where $\Delta=\nabla^a \nabla_a$, the scalar $\chi$ is defined up to a constant, 
the vector field $Z^{(n)}_a$ is divergence-free and
$\pi^{(n)}_{ab}$ is both transverse and tracefree (i.e. $\nabla^a\pi^{(n)}_{ab}=\pi^{(n)a}_{~a} =0$).  
\end{theorem}
There are, however, too many degrees of freedom in the above decomposition, 
and one can set some of the free perturbations functions to zero thus choosing a certain {\em gauge}. For convenience, we consider the {\em synchronous gauge} which, in this case, corresponds to taking 
$$\nabla_a\Phi^{(n)}=\psi^{(n)}=W^{(n)}_a=0.$$
However, we shall prove our main results in a gauge independent manner.

We note that, in this framework, the initial perturbations (\ref{bn1})-(\ref{bn3}) are assumed to be small, in the sense that they result from a Taylor series expansion for each order $O(\epsilon^n)$, with $\epsilon\ll 1$ (see e.g. \cite{Stewart, Mata}), and are such that the metric can be written in "almost-RW" form (holding the inequalities (2.58) of \cite{Ellis}). 
We also note that the dependence of the asymptotic dynamics on more general initial conditions cannot be rigorously studied with the present methods. In order to do that, one could use methods from the theory of hyperbolic partial differential equations, such as the ones used in \cite{Rendall, Reula}.   
\section{Evolution of first order perturbations}
The evolution of first order perturbations in this context is a much studied subject (see e.g. \cite{MBT} and references therein). Here we shall just recall the main results which will be needed for the next section. 
\\
From the EFEs with the perturbed metric (\ref{bn1})-(\ref{bn3}) on FLRW flat backgrounds in the synchronous gauge one finds that the evolution equations for the 
first order perturbations are:
\begin{eqnarray}
&&\label{scalar1}
\phi^{(1)''}+\frac{a'}{a}\phi^{(1)'}-\frac{1}{2} a^2\rho_{\scriptscriptstyle \rm B}\phi^{(1)}=0\\
&& \label{vector1}Z^{(1)'}_a+\frac{a'}{a}Z^{(1)}_a=0\\
&&\label{tensor1}\pi^{(1)''}_{ab}+2\frac{a'}{a}\pi^{(1)'}_{ab}-\nabla^2\pi^{(1)}_{ab}=0,
\end{eqnarray}
where the prime denotes differentiation with respect to conformal time. Note that, in the synchronous gauge, the only non-zero perturbation variables are $\phi^{(1)}, Z_{a}^{(1)}, \pi_{ab}^{(1)}, \chi^{(1)}$ but $\phi^{(1)}$ and $\chi^{(1)}$ are related through the momentum constraint equation (\ref{mra1}). 

It is known that the solutions of (\ref{scalar1}) increase in time and approach a positive constant asymptotically. In turn, the solutions of  (\ref{vector1}) decrease in time and asymptote to zero, while the solutions of (\ref{tensor1}) oscilate and decay asymptotically to a constant value in time (an example of this behaviour is depicted in Figure 1).
\begin{figure}
           \includegraphics[width=6cm]{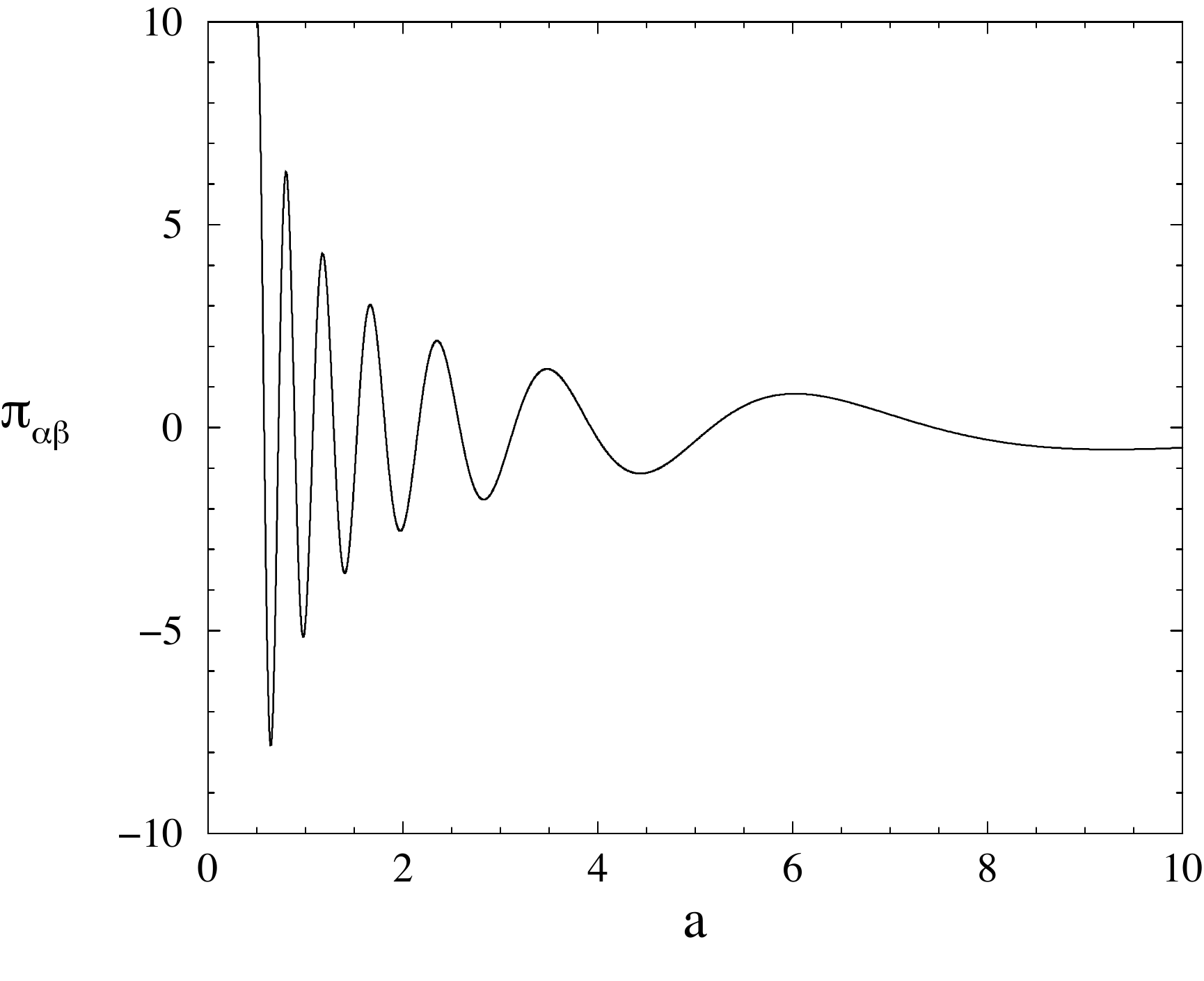} 
           \caption{{Plot obtained from numerical integration of equation (\ref{tensor1}) after Fourier decomposition for $\Lambda=0.001, \rho_0=0.01$ and the Fourier mode $q=1$ (see equations (\ref{mare24}) and (\ref{mare23})).}}
           \end{figure}
One way to show this is to use the asymptotic form for the scale factor which results from the Friedmann equations (see e.g. \cite{MBT}):
\begin{equation}
\label{asymp-a}
a(\tau)=-\sqrt{\frac{3}{\Lambda}}\frac{1}{\tau}.
\end{equation}
Note that as $a\to +\infty$ one has $\tau\to 0$. So, we are interested in the asymptotic behaviour of the solutions to the perturbation equations as $\tau\to 0$.
Using (\ref{scalar1}) and (\ref{asymp-a}) one gets
\begin{equation}
\label{mare18}
\phi^{(1)}(\tau,{\bf x})=C_1({\bf x}) \tau Y\left(\frac{2}{3},\frac{2}{3}A\tau^{3/2}\right)+ C_2({\bf x})\tau 
 J\left(\frac{2}{3},\frac{2}{3}A\tau^{3/2}\right),
\end{equation}
where $A^2=\frac{1}{2}\rho_0\sqrt{\frac{\Lambda}{3}}$, 
$C_1$ and $C_2$ are $C^2$-functions which depend on 
the initial data and
$J$ and $Y$ are the Bessel functions of the first and second
kind, respectively. One can also find a formal solution to (\ref{tensor1})
which can be written in terms of an integral over the wave vector space
\begin{equation}
\label{mare24}
\pi^{(1)}_{ab}(\tau,{\bf x})=\int^\infty_0\pi^{(1)}_{ab}(q,\tau)e^{-i{\bf q}{\bf x}}d{\bf q},
\end{equation} 
with 
\begin{equation}
\label{mare23}
\pi^{(1)}_{ab}(\tau,q)=K_{ab}(q)(\sin{(q\tau)}-q\tau\cos{(q\tau)})+F_{ab}(q)(\cos{(q\tau)}+q\tau\sin{(q\tau)}),
\end{equation}
where $K_{ab}$ and $F_{ab}$ are functions (of the Fourier mode $q$) which depend on the initial data, but are otherwise arbitrary. Finally, from (\ref{vector1}) one finds
\begin{equation}
\label{marealta}
Z^{(1)}_a=\frac{C({\bf x})}{a},
\end{equation}
where $C({\bf x})$ depends on the initial conditions.
So, from (\ref{mare24}), ({\ref{mare23}) and (\ref{marealta}) one gets that the linear perturbations $\phi^{(1)}, \pi^{(1)}_{ab}$ and $Z^{(1)}_a$ tend asymptotically in time to constant values. This corresponds to a local asymptotic approach to the De-Sitter solution. In order to see this explicitly one can use the Boucher-Gibbons \cite{Boucher-Gibbons} local coordinate change to transform the resulting asymptotic form of (\ref{pert-metric}) into the De-Sitter metric (see e.g. \cite{MBT}). This enables to show:
\begin{proposition} 
\label{prop1}
For $n=1$, the metric (\ref{pert-metric}), satisfying the Einstein Field Equations, approaches the De-Sitter solution locally asymptotically in time. 
\end{proposition}
\section{Evolution of second order perturbations and non-linear asymptotic stability of De-Sitter}
One can carry over to second order the first order analysis. 
In this section, we present the evolution and constraint equations for second order perturbations which 
include scalar, tensor and vector modes at first and second orders in presence of a non-zero 
cosmological constant ($\Lambda\ne 0$). These equations are derived in the synchronous gauge from (\ref{evolutiona})-(\ref{energy}) and generalise the equations obtained in
\cite{Mata} for the case of $\Lambda=0$ and zero vector perturbation modes. For convenience, we do not split $\chi_{ab}^{(2)}$ into its vector and tensor parts as before (recall equation (\ref{decomp})) and we write the perturbations in the 3-metric
as $g^{(1)}_{ab}=a^2(\tau)\gamma^{(1)}_{ab}$ and  $g^{(2)}_{ab}=a^2(\tau)\gamma^{(2)}_{ab}$. 
\\\\
{\bf Raychaudhuri equation}
\begin{eqnarray}
\label{lp0}
&&{\phi_{\scriptscriptstyle}^{(2)}}''+
{a' \over a} {\phi_{\scriptscriptstyle}^{(2)}}' - {1 \over 2} \rho_{\scriptscriptstyle \rm B} a^2
\phi_{\scriptscriptstyle}^{(2)} = - {1 \over 6} {\gamma^{(1)ab}}'
\biggl( {\gamma^{(1)}_{{\scriptscriptstyle}ab}}'
+ 2{a'  \over a} \gamma^{(1)}_{{\scriptscriptstyle}ab} \biggr)
- \frac{1}{3}\gamma^{(1)ab}\gamma^{(1)''}_{ab}-\\
&&~~~-{1 \over 6} \rho_{\scriptscriptstyle \rm B} a^2 \biggl[ - {1 \over 4}
\biggl(\gamma^{(1)a}_{{\scriptscriptstyle}a}
- \gamma^{(1)~a}_{{\scriptscriptstyle}0a} \biggr)^2
- {1 \over 2}
\biggl(\gamma^{(1)ab}_{\scriptscriptstyle}
\gamma^{(1)}_{{\scriptscriptstyle}ab} -
\gamma^{(1)ab}_{{\scriptscriptstyle}0}
\gamma^{(1)}_{{\scriptscriptstyle}0ab} \biggr)
+ \delta_0 \biggl(\gamma^{(1)a}_{{\scriptscriptstyle}a} -
\gamma^{(1)~a}_{{\scriptscriptstyle}0a} \biggr) \biggr]\equiv N^{(2)}_1. \nonumber
\end{eqnarray}
{\bf Energy constraint}
\begin{eqnarray}
\label{lp1}
&&{a' \over a} {\phi^{(2)}_{\scriptscriptstyle}}'- {1 \over 3}
\nabla^2 \phi^{(2)}_{\scriptscriptstyle} +\frac{1}{2}\rho_{\scriptscriptstyle \rm B} a^2
\phi^{(2)}_{\scriptscriptstyle} - {1 \over 12}
\chi^{(2)ab}_{{\scriptscriptstyle}~~~,ab}
= - \frac{1}{3}{a' \over a} \gamma^{(1)ab}_{\scriptscriptstyle}
{\gamma^{(1)}_{{\scriptscriptstyle}ab}}' -\nonumber\\
&&~~~~+{1 \over 24} \biggl( {\gamma^{(1)ab}_{\scriptscriptstyle}}'
{\gamma^{(1)}_{{\scriptscriptstyle}ab}}' -
{\gamma^{(1)a}_{{\scriptscriptstyle}a}}'
{\gamma^{(1)b}_{{\scriptscriptstyle}b}}'\biggr)
+
{1 \over 6} \biggl[ \gamma^{(1)ab}_{\scriptscriptstyle}
 \biggl( \nabla^2 \gamma^{(1)}_{{\scriptscriptstyle}ab}
+ \gamma^{(1)d}_{{\scriptscriptstyle}d~,ab} -
2 \gamma^{(1)d}_{{\scriptscriptstyle}a~,bd} \biggr) +\nonumber\\
&&
~~~~+\gamma^{(1)d a}_{{\scriptscriptstyle}~~~,d}
\biggl( \gamma^{(1)b}_{{\scriptscriptstyle}b~,a} -
\gamma^{(1)b}_{{\scriptscriptstyle}a~,b} \biggr)
+ {3 \over 4} \gamma^{(1)ab}_{{\scriptscriptstyle}~~~,d}
\gamma^{(1),d}_{{\scriptscriptstyle}ab}
- {1 \over 2} \gamma^{(1)ab}_{{\scriptscriptstyle}~~~,d}
\gamma^{(1)d}_{{\scriptscriptstyle}a~~,b}
- {1 \over 4} \gamma^{(1)a,d}_{{\scriptscriptstyle}a}
\gamma^{(1)b}_{{\scriptscriptstyle}b~,d} \biggr]
\\
&&~~~~+ \frac{1}{6}\rho_{\scriptscriptstyle \rm B} a^2 \biggl[ -
{1 \over 4} \biggl(\gamma^{(1)a}_{{\scriptscriptstyle}a}
- \gamma^{(1)~a}_{{\scriptscriptstyle}0a} \biggr)^2 - {1 \over 2}
\biggl( \gamma^{(1)ab}_{\scriptscriptstyle}
\gamma^{(1)}_{{\scriptscriptstyle}ab} -
\gamma^{(1)ab}_{{\scriptscriptstyle}0}
\gamma^{(1)}_{{\scriptscriptstyle}0ab} \biggr)
+ \delta_0 \biggl(\gamma^{(1)a}_{{\scriptscriptstyle}a} -
\gamma^{(1)~a}_{{\scriptscriptstyle}0a} \biggr) \biggr]\equiv N^{(2)}_2\nonumber
\end{eqnarray}
{\bf Momentum constraint}
\begin{equation}
\label{lp2}
2 {\phi^{(2)}_{{\scriptscriptstyle},b}}' + {1 \over 2}
{\chi^{(2)a}_{{\scriptscriptstyle}b~,a}}' =
\gamma^{(1)ad}_{\scriptscriptstyle}
\biggl( {\gamma^{(1)}_{{\scriptscriptstyle}bd,a}}' -
{\gamma^{(1)}_{{\scriptscriptstyle}ad,b}}' \biggr)
+ \gamma^{(1)ad}_{{\scriptscriptstyle}~~~~,a}
{\gamma^{(1)}_{{\scriptscriptstyle}bd}}'
- {1 \over 2} \gamma^{(1)ad}_{{\scriptscriptstyle}~~~~,b}
{\gamma^{(1)}_{{\scriptscriptstyle}ad}}'
- {1 \over 2} \gamma^{(1)a}_{{\scriptscriptstyle}a~,d}
{\gamma^{(1)d}_{{\scriptscriptstyle}b}}'\equiv N^{(2)}_{3b}
\end{equation}
{\bf Evolution equation}
\begin{eqnarray}
\label{lp3}
&&-\biggl({\phi^{(2)}_{\scriptscriptstyle}}'' + 2{a' \over a}
{\phi^{(2)}_{\scriptscriptstyle}}' \biggr) \delta^a_{~b} +
{1 \over 2} \biggl( {\chi^{(2)a}_{{\scriptscriptstyle}b}}'' +
2{a' \over a} {\chi^{(2)a}_{{\scriptscriptstyle}b}}' \biggr) +
\phi^{(2),a}_{{\scriptscriptstyle},b} -{1 \over 4}
\chi^{(2)dn}_{{\scriptscriptstyle}~~~,d n} \delta^a_{~b} +\nonumber\\
&&~~~~~~+{1 \over 2} \chi^{(2)da}_{{\scriptscriptstyle}~~~,d b} +
{1 \over 2} \chi^{(2)d,a}_{{\scriptscriptstyle}b~~~,d} - {1 \over 2}
\nabla^2
\chi^{(2)a}_{{\scriptscriptstyle}b}={\gamma^{(1)ad}_{\scriptscriptstyle}}'
{\gamma^{(1)}_{{\scriptscriptstyle}db}}' - {1 \over 2}
{\gamma^{(1)d}_{{\scriptscriptstyle}d}}'
{\gamma^{(1)a}_{{\scriptscriptstyle}b}}' + \nonumber\\
&&~~~~~~+{1 \over 8}
\biggl[\biggl({\gamma^{(1)d}_{{\scriptscriptstyle}d}}'\biggr)^2 -
{\gamma^{(1)d}_{{\scriptscriptstyle}n}}'
{\gamma^{(1)n}_{{\scriptscriptstyle}d}}' \biggr] \delta^a_{~b}
- {1 \over 2} \biggl[ - \gamma^{(1)a}_{{\scriptscriptstyle}b}
\biggl( \gamma^{(1)d,n}_{{\scriptscriptstyle}n~~~,d}
- \nabla^2 \gamma^{(1)d}_{{\scriptscriptstyle}d} \biggr)
\\
&&~~~~~~+ 2 \gamma^{(1)dn}_{\scriptscriptstyle}
\biggl( \gamma^{(1)a}_{{\scriptscriptstyle}b~,dn} +
\gamma^{(1)~,a}_{{\scriptscriptstyle}dn~,b}
- \gamma^{(1)a}_{{\scriptscriptstyle}n~,bd}
- \gamma^{(1)~,a}_{{\scriptscriptstyle}n b~,d}  \biggr)
+ 2 \gamma^{(1)dn}_{{\scriptscriptstyle}~~~,d}
\biggl( \gamma^{(1)a}_{{\scriptscriptstyle}b~,n}
- \gamma^{(1)a}_{{\scriptscriptstyle}n~,b} -
\gamma^{(1),a}_{{\scriptscriptstyle}bn} \biggr)
\nonumber\\
&&~~~~~~
+ 2 \gamma^{(1)ea}_{{\scriptscriptstyle}~~~,n}
\gamma^{(1),n}_{{\scriptscriptstyle}be}
- 2 \gamma^{(1)ea}_{{\scriptscriptstyle}~~~,n}
\gamma^{(1)n}_{{\scriptscriptstyle}b~,e}
+ \gamma^{(1)en}_{{\scriptscriptstyle}~~~,b}
\gamma^{(1)~,a}_{{\scriptscriptstyle}en}
+ \gamma^{(1)n}_{{\scriptscriptstyle}n~,e}
\biggl( \gamma^{(1)ea}_{{\scriptscriptstyle}~~~,b} +
\gamma^{(1)e,a}_{{\scriptscriptstyle}b} -
\gamma^{(1)a,e}_{{\scriptscriptstyle}b} \biggr)
\nonumber\\
&&~~~~~~
- \gamma^{(1)dn}_{\scriptscriptstyle}
\biggl( \nabla^2 \gamma^{(1)}_{{\scriptscriptstyle}dn}
+ \gamma^{(1)m}_{{\scriptscriptstyle}m~,dn} -
2 \gamma^{(1)m}_{{\scriptscriptstyle}d~~,mn} \biggr)
\delta^a_{~b}
- \gamma^{(1)n d}_{{\scriptscriptstyle}~~~,n}
\biggl( \gamma^{(1)m}_{{\scriptscriptstyle}m~,d} -
\gamma^{(1)m}_{{\scriptscriptstyle}d~,m}
\biggr) \delta^a_{~b}
\nonumber\\
&&~~~~~~
- {3 \over 4}
\gamma^{(1)dn}_{{\scriptscriptstyle}~~~,m}
\gamma^{(1),m }_{{\scriptscriptstyle}dn}
\delta^a_{~b}
+ {1 \over 2} \gamma^{(1)dn}_{{\scriptscriptstyle}~~~,m}
\gamma^{(1)m }_{{\scriptscriptstyle}d~~,n}
\delta^a_{~b} + {1 \over 4} \gamma^{(1)d,m}_{{\scriptscriptstyle}d}
\gamma^{(1)n}_{{\scriptscriptstyle}n~,m}
\delta^a_{~b} \biggr]\equiv N_{4b}^{(2)a}.\nonumber
\end{eqnarray}
where $\delta=(\rho-\rho_{\scriptscriptstyle \rm B})/\rho_{\scriptscriptstyle \rm B}$, the subindex $0$ denotes evaluation at an initial time
$\tau_0$ and the commas denote $\nabla$ derivatives.

Notice that, while at first order the perturbation equations for the different modes (scalar, vector and tensor) decouple from each other (recall equations (\ref{scalar1})-(\ref{tensor1})), at second order the perturbation equations (and, therefore, the perturbation modes) are coupled. So, for example, products of first order scalar-tensor, scalar-scalar or scalar-vector modes will, in general, source all second order modes. 

Notice also that the terms $N^{(2)}$ on the r.h.s. of the four equations are of the form $O(1)O(1)$, i.e. these terms result from couplings of first order perturbations. So, from (\ref{mare18})-(\ref{marealta}) we get that all terms $N^{(2)}$ approach constant values asymptotically in time. This will be crucial ahead to find, from (\ref{lp0})-(\ref{lp3}), that all $Z^{(2)}_a$, $\pi^{(2)}_{ab}$ and $\phi^{(2)}$ also tend asymptotically to constants in time (an example of this behaviour is depicted in Figure 2). 

In detail, using the asymptotic expressions (\ref{asymp-a})-(\ref{marealta}) in equation (\ref{lp0}) we can integrate to get the asymptotic estimate
\begin{equation}
\label{mare38}
\phi^{(2)}(\tau,{\bf x})=\frac{L({\bf x})}{A^2}+ O(\tau^2),
\end{equation}
where $L({\bf x})$ is a constant function in time whose expression can be obtained from the asymptotic form of $N^{(2)}_1$. In turn, from integration of (\ref{lp1}) and (\ref{lp2}) we obtain asymptotic conditions in the form 
\begin{eqnarray}
\nabla_\a\chi^{(2)\a}_{\b}(\tau,{\bf x})&=& Q_\b({\bf x})+O(\tau^2)\nonumber\\
\label{wate}
\nabla_\a\nabla_\b\chi^{(2)\a\b}(\tau,{\bf x})&=& G({\bf x})+O(\tau^2)
\end{eqnarray}
where, again, $Q_\b$ and $G$ are constant functions in time. 
Also, substituting the expressions (\ref{asymp-a})-(\ref{marealta}) in equation (\ref{lp3}), the latter can formally be integrated to give
\begin{equation}
\label{mare43}
\chi^{(2)}_{\a\b}(\tau,{\bf x})=\int^\infty_0\chi^{(2)}_{\a\b}(\tau,k)e^{-i{\bf k}{\bf x}}d{\bf k},
\end{equation}
with\begin{equation}
\label{mare42}
\chi^{(2)}_{\a\b}(\tau,k)=\frac{E_{\a\b}(k)}{k}+O(\tau^2)
\end{equation}
where $E$ is a function of the Fourier modes $k$ which depends on the initial conditions.

From (\ref{mare38})-(\ref{mare42}) it is then clear that both $\phi^{(2)}$ and $\chi_{ab}^{(2)}$ tend to constant values in time asymptotically. Finally, to see that $Z^{(2)}_a$ individually tends to constant values we substitute $\chi_{ab}^{(2)}=2\nabla_{(a}Z^{(2)}_{b)}+\pi^{(2)}_{ab}$ into equation (\ref{lp3}).  Given
$\nabla^a\pi^{(2)}_{ab}=0$, then there will be no $\pi_{ab}^{(2)}$ terms in (\ref{lp3}). Now, since the first order terms (on the r.h.s.) and the terms in $\phi^{(2)}$ (on the l.h.s.) are already known to asymptote to constants, then the remaining term in $Z_a^{(2)}$ will also tend to a constant value in time. We can use this in equation (\ref{lp3}) to show that $\pi^{(2)}_{ab}$ also tends to constants. 

Therefore, we conclude that all second perturbations $\phi^{(2)}, \pi^{(2)}_{ab}$ and $Z^{(2)}_a$ tend asymptotically in time to constant values. We note that the density perturbations can be written in terms of the metric perturbations and the expressions that relate these quantities have been derived in \cite{Mata}. In the present case, since the metric perturbations asymptote to constants in time, so do the density perturbations.
As before, one can now apply the Boucher-Gibbons \cite{Boucher-Gibbons} local coordinate change to explicitly transform the resulting asymptotic form of (\ref{pert-metric}) into the De-Sitter metric. 

One can also see this in a coordinate and gauge invariant way: Using (\ref{mare38})-(\ref{mare42}) we can show
that at second order, asymptotically, 
\begin{eqnarray}
\label{mare49}
&& \theta^2\to \Lambda, ~~~ H_{\a\b}\to 0,~~~R\to 4\Lambda, \\
\label{mare50}
&& \sigma^2 \to 0,~~~E_{\a\b}\to 0,~~~R^*\to 0,   
\end{eqnarray}
where $\sigma$ is the shear scalar, $H_{\a\b}$ and $E_{\a\b}$ are the magnetic and electric parts of the Weyl tensor, respectively. The asymptotic behaviour (\ref{mare49})-(\ref{mare50}), together with the fact that $\dot u_\a=0$ and $\omega_{\a\b}\to 0$, where $\omega_{\a\b}$ is the vorticity, caracterizes gauge invariantly the approach to the De-Sitter solution (see e.g. \cite{Ellis}). 

This result generalises the result of Proposition \ref{prop1} and previously known second order results  \cite{MBT} by including, at second order, sources of first order vector and tensor perturbation modes as well as pure second order vector modes: 
\begin{proposition} 
\label{prop2}
For $n=2$, the metric (\ref{pert-metric}), satisfying the Einstein Field Equations, approaches the De-Sitter solution locally asymptotically in time. 
\end{proposition}
\begin{figure}
           \includegraphics[width=6cm]{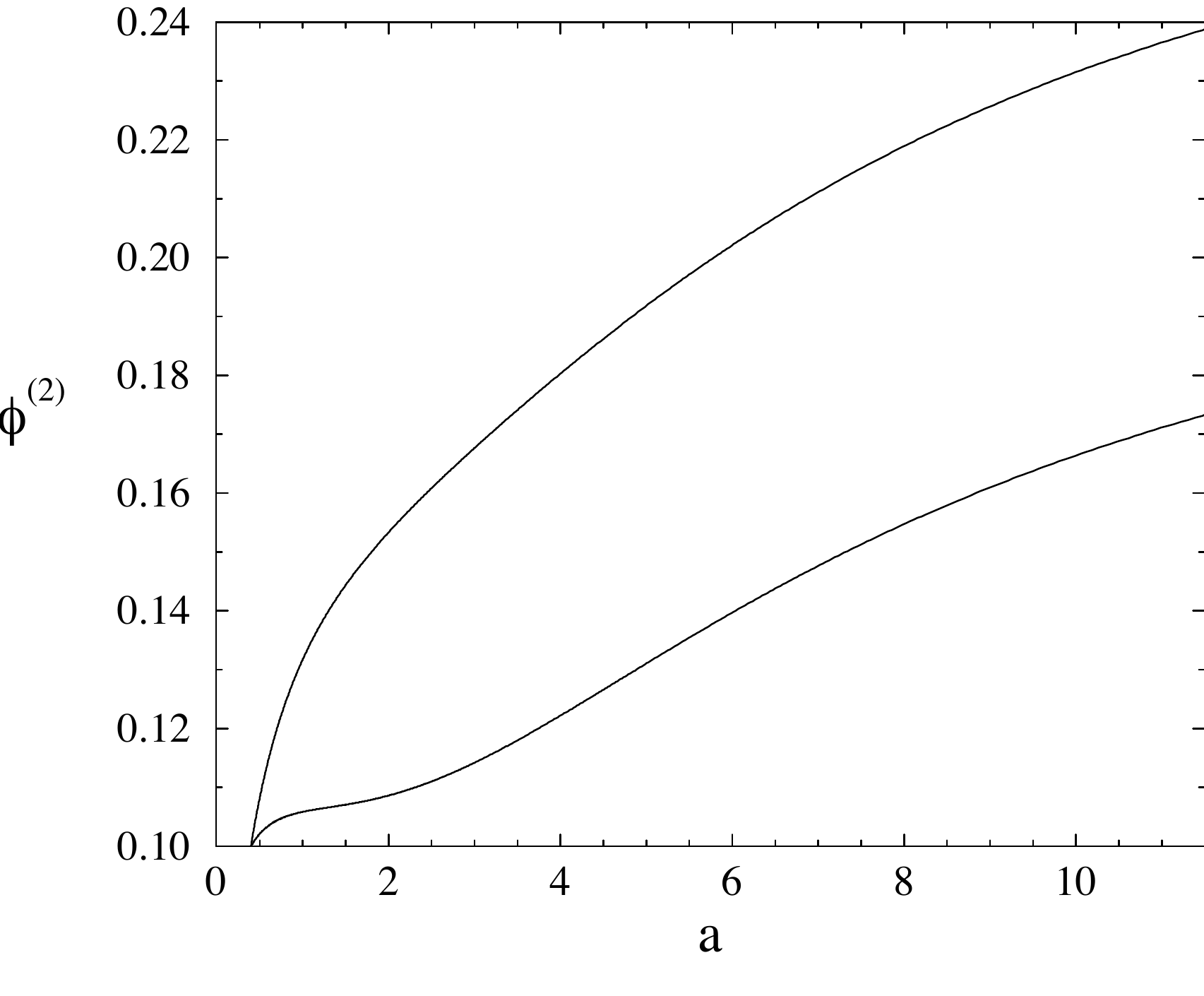} 
           \caption{{Two plots obtained from numerical integration of equation (\ref{lp0}) for the 
           evolution of second order scalar perturbations $\phi^{(2)}$ for a $k=0$ FLRW dust background with 
           $\Lambda=0.001$ and $\rho_0=0.01$.}}
           \end{figure}
Consider now $n$th order perturbations in (\ref{pert-metric}). It is clear that, as before, the sources of the EFEs for the $n$th order perturbations are products of lower  order terms such as $O(n-1)O(1), O(n-2)O(2)$ etc. For example:
\begin{equation}
{\phi_{\scriptscriptstyle}^{(n)}}''+
{a' \over a} {\phi_{\scriptscriptstyle}^{(n)}}' - {1 \over 2} \rho_{\scriptscriptstyle \rm B} a^2
\phi_{\scriptscriptstyle}^{(n)}=N_1^{(n)},
\end{equation}
where $N_1^{(n)}$ is a sum of products of lower order terms. By proving that the lower order terms tend to constants asymptotically in time, one can proceed as before, order by order, and show, by induction, the following result: 
\begin{theorem}
At each order $n$, the truncated metric (\ref{pert-metric}), satisfying Einstein Field Equations, is convergent and approaches the De-Sitter solution locally asymptotically in time. 
\end{theorem}
\vspace{0.5cm}
{\bf Acknowledgements:}
\\\\
The author is funded by FCT (Portugal) through project\\ PTDC/MAT/108921/2008, thanks Centro de Matem\'atica, University of Minho for support and Dep. Matem\'atica, IST, Lisbon, where this work has been completed.

\end{document}